\documentstyle[aps,12pt]{revtex}
\begin{document}
\begin{titlepage}
\def\today{\ifcase\month\or
        January\or February\or March\or April\or May\or June\or
        July\or August\or September\or October\or November\or December\fi,
  \number\year}
\rightline{hep-th/yymmddd}
\rightline{July, 1996}
\vskip 1cm
\centerline{\Large \bf Integrable Structure in  
SUSY Gauge Theories,} 
\centerline{\Large \bf and String Duality\footnote{
Talks given at the {\it Inauguration Conference of APCTP},
June 4-10, 1996, Seoul, Korea,
and at the {\it Argonne Duality Institute}, June 27-July 12, 1996, Argonne Nat. Lab., USA.}}
\vskip 2cm
\centerline{
{\sc Soonkeon Nam} }
\vskip 1cm
\centerline {{\it Department of Physics and}} 
\centerline {{\it Research Institute for Basic Sciences,}}
\centerline {{\it Kyung Hee University, Seoul 130-701, Korea}}
\centerline {{\tt nam@nms.kyunghee.ac.kr}}
\vskip 1cm
\centerline{\large \sc Abstract}
\vskip 0.2in
There is a close relation between duality in $N=2$ SUSY gauge theories
and integrable models.
In particular, the quantum moduli space of vacua of $N=2$ SUSY 
$SU(3)$ gauge theories coupled to two flavors of massless quarks
in the fundamental representation can be related to the spectral curve
of the Goryachev-Chaplygin top.
Generalizing this to the cases with {\it massive} quarks,
and $N_f = 0,1,2$, we find a corresponding integrable system in
seven dimensional phase space where a hyperelliptic curve appears in the Painlev\'e
test.  To understand the stringy origin of the integrability of these theories
we obtain exact nonperturbative point particle limit 
of type II string compactified on a Calabi-Yau manifold, which gives
the hyperelliptic curve of $SU(2)$ QCD with $N_f =1$ hypermultiplet.

\end{titlepage}
\newpage

\def\beq{\begin{equation}}
\def\eeq{\end{equation}}
\def\bea{\begin{eqnarray}}
\def\eea{\end{eqnarray}}
\renewcommand{\arraystretch}{1.5}
\def\ba{\begin{array}}
\def\ea{\end{array}}
\def\bce{\begin{center}}
\def\ece{\end{center}}
\def\nn{\noindent}
\def\nonu{\nonumber}
\def\pbx{\partial_x}


\def\ptl{\partial}
\def\al{\alpha}
\def\be{\beta}
\def\ga{\gamma} 
\def\Ga{\Gamma}
\def\de{\delta} \def\De{\Delta}
\def\ep{\epsilon}
\def\vep{\varepsilon}
\def\ze{\zeta}
\def\et{\eta}
\def\th{\theta} \def\Th{\Theta}
\def\vth{\vartheta}
\def\io{\iota}
\def\ka{\kappa}
\def\la{\lambda} 
\def\La{\Lambda}
\def\rh{\rho}
\def\si{\sigma} \def\Si{\Sigma}
\def\ta{\tau}
\def\up{\upsilon} 
\def\Up{\Upsilon}
\def\ph{\phi} 
\def\Ph{\Phi}
\def\vph{\varphi}
\def\ch{\chi}
\def\ps{\psi} 
\def\Ps{\Psi}
\def\om{\omega} 
\def\Om{\Omega}

\def\lbr{\left(}
\def\rbr{\right)}
\def\half{\frac{1}{2}}
\def\CVO#1#2#3{\!\left( \matrix{ #1 \cr #2 \ #3 \cr} \right)\!}

\def\vol#1{{\bf #1}}
\def\nupha#1{Nucl. Phys. \vol{#1} }
\def\phlta#1{Phys. Lett. \vol{#1} }
\def\phyrv#1{Phys. Rev. \vol{#1} }
\def\PRL#1{Phys. Rev. Lett \vol{#1} }
\def\prs#1{Proc. Roc. Soc. \vol{#1} }
\def\PTP#1{Prog. Theo. Phys. \vol{#1} }
\def\SJNP#1{Sov. J. Nucl. Phys. \vol{#1} }
\def\TMP#1{Theor. Math. Phys. \vol{#1} }
\def\ANNPHY#1{Annals of Phys. \vol{#1} }
\def\PNAS#1{Proc. Natl. Acad. Sci. USA \vol{#1} }
\def\CMP#1{Comm. Math. Phys. \vol{#1} }

Last several years we have witnessed very important progress in 
understanding $S$-duality of $N=2$ SUSY gauge 
theories\cite{SW}. The low energy description of these theories
can be encoded by Riemann surfaces and 
the integrals of meromorphic one differentials over the periods of them.
Exact effective actions of these theories 
can be described by holomorphic functions,
so-called prepotentials. With these we can probe the strong coupling limit 
of the theories.
There are other type of theories where this structure on 
Riemann surfaces plays a crucial role. These are the integrable models in
two dimensions\cite{DKN}. Among the methods of solving integrable models, in
inverse scattering method we obtain the solitons solutions as potentials
of a quantum mechanics problem, given the scattering data.
The spectral parameter plays the role of the energy.
If we consider the periodic soliton solutions, then the spectral parameter
develops forbidden zones, just as we are familar in solid state physics.
Analytic continuation of the spectral parameter with the forbidden zones
 gives us the Riemann surface with genus $g>0$.
By now there are many works which connect these low-energy effective
theories with known integrable systems.
To relate effective quantum field theories with integrable systems, 
one needs averaging over fast oscillations, i.e. Whitham averaging.
It was analyzed that the periods of the modulated Whitham solution
of periodic Toda lattice give rise to the mass
spectrum in the BPS saturated states\cite{GKMMM,NT}.
Furthermore this framework of Whitham 
dynamics for the Toda lattice
was generalized to other gauge groups\cite{MW}. 
For the case of $SU(N_{c})$ gauge theory with a single
hypermultiplet in the {\it adjoint} representation, 
the corresponding integrable system was found\cite{DW} 
and was recognized to be the elliptic spin Calogero model\cite{Mar}. 
This connection was developed in Ref.\cite{IM} by identifying the 
coupling constant of Calogero system with the mass of a 
hypermultiplet in the adjoint
representation, starting from the Lax operator for the Calogero model and
calculating the spectral curve explicitly.
The integrable system related to 
gauge theories coupled to {\it massive } hypermultiplets 
in the {\it fundamental } representation was discussed\cite{AN,RUS,Brz},
and we will be explaining some of the results of Ref.\cite{AN}. 

Motivated by the works in SUSY gauge theories, the duality
really blossomed in the context of string theories\cite{JP}.
Among these, the $N=2$ type II/heterotic duality in four dimensions 
has been proposed in \cite{KV} and further studied in many subsequent
papers.  In fact, it was extended to the $F$-theory/heterotic duality\cite{MV}
in eight dimensions
where the heterotic strings compactified on $T^2$ is dual to 
$F$-theory compactified on $K3$ which admits an elliptic fibration.
Further compactification in six dimensions leads to
the duality between $F$-theories compactified on Calabi Yau(CY) manifolds and
heterotic strings on $K3$.
Among the many ways to check the consistency on this duality 
one can consider the point like limit of four dimensional
$N=2$ SUSY compactifications of heterotic strings, and
see the resulting gauge theory\cite{KKLMV}, which reproduces the
exact field theory results of Ref.\cite{SW}.
Additional question would be whether one can get also matter
from the point like limit of the string theory compactification.
We would like to see whether the $N=2$ SUSY QCD is embedded in this
compactification of string theory\cite{KHU}.

In this talk, we first consider the $N =2$ SUSY $SU(N_c)$ gauge
theories with $N_c$ colors and $N_f$ flavors. The field content of the 
theories
consists, in terms of $N=1$ superfields, a vector multiplet $W_\al$,
a chiral multiplet $\Ph$, and two chiral superfields $Q^i_a$ and
$\tilde{Q}_{ia}$ where $i= 1, \cdots, N_f$ and $a =1, \cdots, N_c$.
The superpotential reads,
\beq
 W = \sqrt{2} \tilde{Q}_i \Ph Q ^i + \sum_{i=1}^{N_f} m_i \tilde{Q}_i Q^i,
\eeq
where $m_i$'s are the bare quark masses and color indices are suppressed.
The curve representing the moduli space with
$N_{f} < N_{c}$ case is as follows\cite{HO}:
\bea
y^2=(x^{N_{c}}-\sum_{i=2}^{N_{c}} u_{i} x^{N_{c}-i})^2-
\Lambda_{N_{f}}^{2N_{c}-N_{f}} \prod_{i=1}^{N_{f}} (x+m_{i}),
\label{eq:curve}
\eea
where the moduli $u_{i}$'s are the vacuum expectation 
values of a scalar field
of the $N=2$ chiral multiplet, and $m_i$'s are the bare quark masses. 
It turns out that from the point of view of integrable theory, 
$u_{i}$'s correspond to the integrals of motion. 
The second term in Eq.(\ref{eq:curve}) is due to the instanton corrections.
For the $N_c \leq N_f < 2 N_c$ case, 
the correction due to matter is different and the curve
is given in Ref.\cite{HO}.
By inspection we see that the case of $N_f = 0$ corresponds to the periodic 
Toda lattice with $N_c$-particles,
after an appropriate rescaling of the variables\cite{GKMMM}.
In general the following type of hyperelliptic curve appears 
\beq
y^2 = P_{N_c} (x)^2 - Q_m (x),
\eeq
where $P_n(x)$ and $Q_m (x)$ are polynomials of order $n$ and $m$.
It is natural to ask which integrable theories have such spectral curves.
The form is indicative of Riemann surfaces with punctures as well as genus.
We will start with the known case of $y^2 = P_3 (x)^2 - a x^2$ ($a$ is a constant) 
which corresponds to the so called Goryachev-Chaplygin (GC) top.
It was noted in Ref.\cite{Marsha} that there exists such a connection.

Let us review the classical mechanics of rotation of a heavy rigid body 
around a fixed point, which is described by the following Hamiltonian:
\bea
H(M,p)=\frac{M_{1}^2}{2I_{1}} +\frac{M_{2}^2}{2I_{2}}+
\frac{M_{3}^2}{2I_{3}}+\gamma_{1} p_{1}+\gamma_{2} p_{2}+
\gamma_{3} p_{3}.
\label{eq:hamil}
\eea
The phase space of this system is six dimensional: 
$M_i$'s are the components of the angular momentum and $p_i$'s are
the linear momenta.
$I_{i}$'s are the principal
moments of inertia of the body and $\gamma_{i}$'s
are the coordinates of the center of mass. 
There are four known integrable cases for the Hamiltonian in 
Eq.(\ref{eq:hamil}). 
In all these cases there is always one obvious integral of motion, the energy. 
It is necessary to get one extra integral independent of 
the energy for complete integrability 
according to Liouville's theorem\cite{DKN}.

Apart from the better known cases of Euler's and Lagrange's tops,
we have following two other cases:
i) Kowalewski's case: ($ I_{1}=I_{2}=2I_{3}, \gamma_{3}=0$. )
The extra integral can be found by the Painlev\'{e} test or the
Kowalewski's asymptotic
method. Here the symmetry group is $SO(3,2)$.       
ii) Goryachev-Chaplygin's case: 
($I_{1}=I_{2}=4I_{3}, \gamma_{3}=0$) We need 
$M_{1} p_{1}+M_{2} p_{2}+M_{3} p_{3}=0$, which leads to a new integral of 
motion
together with the Hamiltonian $H$ and the GC integral $G$\cite{Koz}.
The Lax operator for the GC top  is given as follows\cite{BK}: 
\bea
L(z)=
\left( \begin{array}{ccc}
0 & -i p_3/z & M_2-i M_{1} \\
i p_{3}/{z} & 2 i M_{3} & -2 i z +(p_{2}-i p_{1})/z  \\
-M_{2}-i M_{1} & 2 i z+(p_{2}+ i p_{1})/z & -2 i M_{3}  \\
\end{array} \right).
\eea
This Lax operator depends on the phase space variables, $M_i, \  p_i$ 
and on the spectral parameter, $z$.
Now it is easy to calculate the spectral curve from the equation
${\rm Det}( L(z)-x I ) =0$, which gives the spectral curve
as follows:
\bea
x^3+2 x H -2 i G-x (4 z^2+\frac{\lambda^2}{z^2}) =0,
\eea
where $H=\frac{1}{2} ( M_{1}^2+M_{2}^2+4 M_{3}^2 )-2 p_{1}$ is the
Hamiltonian, and $G=M_{3} (M_{1}^2+M_{2}^2)+2 M_{1} p_{3}$ is the 
GC integral. 
We also have the following constraints: 
\beq
p_{1}^2+p_{2}^2+p_{3}^2=\lambda^2,\;\;\; {\rm and}\;\;\; M_{1} p_{1}+
M_{2} p_{2}+M_{3} p_{3}=0 . 
\eeq
Now we see that the spectral curve
depends on special combinations of $M_i, p_i$'s, which are nothing but 
the integrals of motion.
By introducing variable $y=x (4 z^2-\frac{\lambda^2}{z^2})$, we thus get
\bea
y^2=(x^3+2 H x-2 i G)^2-16 \lambda^2 x^2,
\label{eq:curve1}
\eea
which are the same as the curve for GC top with some rescalings.
To relate this to the curve of SUSY gauge
theory we make the following simple substitutions:
\beq
H \rightarrow -\frac{1}{2} u_{2}, \;\;\;\;\; 
G \rightarrow -\frac{i}{2} u_{3},
\;\;\;\;\; \lambda^2 \rightarrow \frac{1}{16} \Lambda_{2}^{4}.
\eeq
It is easy to see that Eq.(\ref{eq:curve1}) exactly coincides 
with Eq.(\ref{eq:curve}) for the particular
case of $N_{c}=3, N_{f}=2$ and $m_{1}=m_{2}=0$!

Since we have seen the intimate relation between the GC top and 
the SUSY $SU(3)$ gauge theory with two flavor
massless hypermultiplets, it is natural for us to extend
this to the massive case. For this purpose we need an integrable system
which has both the GC top and the three body 
Toda lattice as particular limits, because the latter corresponds to
pure gauge theory with no matter.
The Hamiltonian system which realizes this is hard to imagine,
but there exists a system of coupled seven nonlinear differential
equations in mathematical literature\cite{BvM}.
This system has the following nonlinear ``equations of motion":
\bea
&&\dot{z_{1}}  =  -8 z_{7}, \ \ \ \ 
\dot{z_{2}}  =  4 z_{5}, \ \ \ \
\dot{z_{3}}  =  2 (z_{4} z_{7}-z_{5} z_{6}), \ \ \ \ 
\dot{z_{4}}  =  4 z_{2} z_{5}-z_{7}, \nonu \\
&&\dot{z_{5}}  =  z_{6}-4 z_{2} z_{4}, \ \ \ \  
\dot{z_{6}}  =  -z_{1} z_{5}+2 z_{2} z_{7}, \ \ \ \
\dot{z_{7}} =  z_{1} z_{4}-2 z_{2} z_{6}-4 z_{3}.
\label{eq:diff}
\eea
There are following five constants of motion of the system:
\bea
&&6 a  =  z_{1}+4 z_{2}^2-8 z_{4}, \ \
2 b  =  z_{1} z_{2}+4 z_{6}, \ \
c    =  z_{4}^2+z_{5}^2+z_{3}, \nonu \\
&&d    =  z_{4} z_{6}+ z_{5} z_{7}+z_{2} z_{3}, \ \
e    =  z_{6}^2+z_{7}^2- z_{1} z_{3}.
\label{eq:integral}
\eea
Although the Lax operator for this system is not readily available, 
we can still apply the asymptotic method due to Kowalewski to this system 
and take $z_i=t^{-n_i} \sum_{j=0}^{\infty} A^{i}_{j} t^{j}$ where
$n_i$'s are positive integers\cite{SE,BvM}. 
Substituting these Laurent expansions
into the system of Eqs.(\ref{eq:diff}) and (\ref{eq:integral}),
one finds $n_i=1$ for $i=1, 2, 3$,
$n_i=2$ for $i=4, 5, 6, 7$ and a relation between the coefficients of 
$A^{i}_{j}$'s.
Then we obtain
the Laurent solutions for this system with seven parameters, 
five of which are from the constants of motion, $a, b, c, d, e$
and two additional ones $x$ and $y$ where they satisfy the equation
for an hyperelliptic curve\cite{BvM}:
\bea
y^2=(2 x^3-3 a x+b)^2 -4 ( 4 c x^2+4 d x+ e).
\label{eq:curve2}
\eea
We clearly see that with the following  substitution this gives the algebraic
curves given in Eq.(\ref{eq:curve}) of $N=2$ SUSY  $SU(3)$  
gauge theories with massive quarks of two flavors of masses $m_1$ and $m_2$:
\beq
y \rightarrow 2 y, \;\;\;\; a \rightarrow \frac{2}{3} u_{2}, 
\;\;\;\;
b \rightarrow -2 u_{3}, \ \ \ \
c \rightarrow \frac{1}{4} \Lambda_{2}^{4}, \;\;\;\; d \rightarrow
\frac{\Lambda_{2}^{4}}{4} (m_{1}+m_{2}), \;\;\;\;
e \rightarrow \Lambda_{2}^{4} m_{1} m_{2}.
\label{eq:eq19}
\eeq
When we consider the case of $c=0$, then this
leads to gauge theory coupled to one massive quark of 
mass $m_1$ or massless one$(N_f=1)$
after similar substitution. 
For the case of $c=d=0$,
the usual periodic Toda lattice
is recovered, and for $d=e=0$ we get back GC top.
So clearly we have a unifying model of two seemingly different systems.

Now let us consider the point like limit of string theories, where
the $N=2$ SUSY QCD is embedded in a compactification of string theory.
We obtain exact nonperturbative point particle limit  of a four
dimensional $N=2$ SUSY compactification of heterotic strings.
Using Heterotic/type II duality, 
we show how $N=2$ SUSY QCD 
with one flavor of massless quark arises
in type II string compactification on Calabi-Yau manifolds.

Such analyses were performed for the following two cases\cite{KV}:
First is the case where the $E_8\times E_8$ heterotic string compactified
on $K3\times T^2$ is dual to the type IIB(or type IIA) 
theory compactified on a CY manifold (or its mirror), which is 
the weighted projective space of weights 1,1,2,2,6\cite{COFKM,HKTY}. 
The point like limit of this model was shown to yield the exact results
of \cite{SW} with pure $N=2$ Yang-Mills theory with gauge group $SU(2)$.
Second  case is  where the weighted projective space has weights 
1,1,2,8,12\cite{HKTY},  the point like limit is known to be the that of pure 
$N=2$ $SU(3)$ Yang-Mills theory.
By going to the conifold locus of the CY manifold and blowing it
up,  one can indeed obtain the algebraic curves for all the cases of  
$SU(n)$ gauge groups\cite{KLMVW}.

In order to relate these gauge theories with {\it matter} 
with string compactification scheme on a CY manifold, 
we look for the known cases where the
explicit forms of the discriminant and the Picard-Fuchs operators of the
CY manifolds have been worked out. 
One of the strong candidate is that of the weighted projective space with
weights 1,1,1,6,9\cite{HKTY,CFKM}.
This is because if we look at the discriminant locus in term of the coordinates
describing the large moduli parameters, 
the singularity structure of this is identical to that of 
$N=2$ $SU(2)$ gauge theory coupled to single $(N_f=1)$ flavor 
in the fundamental representation\cite{KV}. 
The dual is the heterotic $E_8 \times E_8$ string compactificated 
on $K3$ with $SU(2)$ bundles with instanton numbers (13, 11) 
has the hypermultiplet spectrum of
$\frac{9}{2} ( 56,1) + \frac{7}{2} (1,56) + 62(1,1)$.

Let therefore consider the moduli space of the mirror of the weighted projective
space with weights 1,1,1,6,9 CY manifold with Hodge numbers 
$h_{1,1}=2, \;\; h_{2,1}=272$ whose defining polynomial given 
as follows\cite{CFKM}:
\beq
p = x_1^{18} + x_2^{18} + x_3^{18}+x_4^{3}+x_5^{2}
-18\ps x_1 x_2 x_3 x_4 x_5 - 3\ph x_1^6 x_2^6 x_3^6=0.
\eeq
This CY manifold has 2 vector multiplets whose scalar 
expectation values correspond
to $\ps$ and $\ph$ and 273 hypermultiplets including a dilaton field.
It is convenient to introduce the following variables that were used for
the complex moduli of the mirror:
\beq
x = \frac{3\ph}{(18\ps)^6},\ \ \ y = \frac{1}{(3\ph)^3}.
\eeq
The discriminant can be written as\cite{CFKM,HKTY}:
$
\Delta = (1-\bar{x})^3 - \bar{x}^3 \bar{y},
$
where $\bar{x} =2^4 3^3 x , \;\; \bar{y}=3^3 y$.
For weak coupling, $\bar{y} \rightarrow 0$, there exists a triple singularity at $\bar{x}=1$.
The locus on which the CY manifold aquires a conifold point is where $\De = 0$.

In order to go to the point like limit of strings ($\al ' \rightarrow 0$)
we would like to identify $\bar{x} -1$ with the vacuum expectation value of 4D gauge 
theory $u$ upto leading order of  $\alpha'$. In fact, to be dimensionally correct we need
\beq 
\bar{x} = 1 + \al' u + {\cal{O}}(\al'^2)=1+\frac{\ep}{\La_1^2} u +{\cal{O}}(\ep^2),
\eeq
where $\La_1$ is the renormalization scale parameter of the
theory with $N_f=1$.
At the conifold locus, we have
\beq
\bar{y} = \frac{(1-\bar{x})^3}{\bar{x}^3}\frac{1}{u^3}.
\eeq
When we expand for $\ps$ and $\ph$ we get
\beq
\ps = \frac{1}{18} \ep^{-\frac{1}{6}}( 1+ \ep \ps_1 + \cdots) , 
\ \ \  \ph = \frac{1}{3} \ep^{-1} (1+ \ep u+ \cdots),
\eeq
where $\ps_1 $ is independent of $u$.
With the expressions in the defining polynomial, we can now
compare with the the curve of SUSY QCD along the line of Ref.\cite{KLMVW}.
From the requirement that the coefficient of the term linear in $u$ be order of
$\ep$, 
we immediately see that the product of $x_1^6 x_2^6 x_3^6$ should be order of $\ep$.
Taking the following expansion, 
\bea
&&x_1  =  \ep^{\frac{1}{18}} a_1+\cdots, \;\;\; x_2=\ep^{\frac{1}{9}} a_2+\cdots ,\ \ \
x_3  = a_3(1+\ep b_3+\cdots), \nonu \\
&& x_4=a_4(1+\ep b_4+\cdots), \ \ \  
x_5 =  a_5(1+\ep b_5+\cdots),
\eea
and by requiring that $p$ has the following form up to the first power of 
$\ep$, we recover the hyperelliptic curve for $SU(2)$ $N_f=1$ gauge theory: 
\beq
p = \ep \left( 2u - 2x^2 + \hat{z} +\frac{\La_{1}^{3}(x+m)}{\hat{z}}+v^2+w^2\right) + 
{\cal{O}}(\ep^2) ,
\eeq
once we fix the functions of $a_i$'s and $b_i$'s as follows:
\bea
 & &a_1  =  (\hat{z})^{\frac{1}{18}}, \;\;\;
 a_2=\left(\frac{2^3 3 x^2}{\La^3 (x+m)}\right)^{\frac{1}{18}} ,\ \ \
a_3  =\left (-\frac{\La^3 (x+m)}{3 x^2 \hat{z}}\right)^{\frac{1}{18}}, \\ \nonu
& &2 a_5 =  (-2)^{\frac{1}{6}} a_4+
\sqrt{-8+(-2)^{\frac{1}{3}} a_4^2-4 a_4^3+
\frac{4 \La^3 (x+m)}{3 x^2 \hat{z}}}, \\ \nonu
& &b_3  = -\frac{x^2}{6}, \;\;\;
b_4=\frac{y^2}{3 a_4^3}, \ \ \ 
b_5  =  \frac{w^2}{2 a_5^2}, \;\;\; \ps_1=\frac{x^2}{6}-\frac{y^2}{3 a_4^3}-b_5.
\eea
Note that the change of variable $y=\hat{z}-P_2(x)$ gives rise to the explicit form
of the curve given in Eq.(\ref{eq:curve}).

Now we consider the periods.
As is the case of pure SUSY Yang-Mills theory\cite{KLMVW}, $p=0$ differs from 
(\ref{eq:curve}) by quadratic terms.
On the other hand,
the holomorphic 3-form\cite{CFKM} is
\beq
\Om = d \left( \ln \frac{\hat{z}}{\sqrt{Q(x)}
} \right) \wedge\left[\frac{dv\wedge dx}{\frac{\partial p}
{\partial w}}\right] .
\label{eq:3form}
\eeq
In order to integrate $\Om$ over $v$ by following similar
arguments in \cite{KLMVW}, we solve for $w$ from $p=0$.
Plugging this value of $w$ into (\ref{eq:3form}), then the integral over $v$
becomes trivial. This leads to the following result:
\beq
\int_y \Om =dx d \ln \frac{\hat{z}}{\sqrt{Q(x)}}=
d \left( x\;d \ln \frac{\hat{z}}{\sqrt{Q(x)}} \right)
\eeq
Now we see that the integral of $\Om$ on a 3-cycle of 
the CY manifold produces an integral of $d S$ over the cycle of Riemann surface.

Now we conclude with a list of some further works to be done.
i)If one wishes to obtain the
prepotentials which are needed for exact effective action in 
SUSY gauge theory, we should consider quasiclassical $\tau$
fuctions in the context of integrable theory as in the case of pure
gauge theory\cite{NT}.
It would be interesting to find out intimate relation between them
by using the explicit form of Baker-Akiezer 
function for GC top\cite{BK}. 
ii)There are algebraic curves for higher rank cases with generic $N_c$ and 
$N_f$.
The obvious thing to do would be to obtain a `higher' dimensional 
generalization of GC top, at least for the massless cases.
Although there exists multi-dimensional generalization\cite{BRS} of Kowalewski 
top, it is not available for GC top.
Nevertheless, with all the results from
SUSY gauge theories pointing to the existence of higher dimensional
generalizations, it is quite tempting to speculate that there exists 
higher dimensional GC top.
iii) As regards the massive cases, better understanding of the variables,
$z_i$'s$(i=1, \cdots, 7)$ in Eq.(\ref{eq:diff}) are needed as well as
the symmetry of the system. In this respect, the relation to the quadratic
algebra might shed further light in the problem in Refs.\cite{KT,Kul}.
Of course it would be nice to find similar integrable theories for other
gauge theories coupled with real matter. 
iv) String duality and integrability of SUSY gauge models are closely related, but
still needs further systematic investigation, especially when we have matter\cite{KHU,KLMVW}.
Especially it would be nice to compare results from field theory calculations\cite{IY,BF}.

This is based on the works done in collaboration with Changhyun Ahn 
and S. Hyun. We have benefited from the discussions with 
A. Morozov, and N. Warner.
I would like to thank A. Chodos for hospitality at Yale.
This work is supported in part by Ministry of Education (BSRI-96-2442), 
KOSEF-JSPS exchange program, KOSEF 961-0201-001-2,  and by CTP/SNU.
\noindent


\begin{thebibliography}{[00]}
\vskip -0.5cm
\bibitem{SW} N. Seiberg and E. Witten,  Nucl. Phys. {\bf B426} (1994) 19;
     {\it Err.}: ibid. {\bf B430} (1994) 485;
 Nucl. Phys. {\bf  B431} (1994) 484.
\bibitem{DKN} B.A. Dubrovin, I.M. Krichever, and S.P. Novikov, in {\it  
"Dynamical Systems. IV: Symplectic Geometry and Its Applications''}, 
V.I. Arnol'd and S.P. Novikov(Eds.), Springer-Verlag (1990), New York. 
\bibitem{GKMMM} A. Gorsky, et al., Phys. Lett. {\bf 355B} (1995)  466.
\bibitem{NT} T. Nakatsu and K. Takasaki, Mod. Phys. Lett. {\bf A11} (1996) 157.
\bibitem{MW} E. Martinec and N. Warner, Nucl. Phys. {\bf B459} (1996) 97.
\bibitem{DW} R. Donagi and E. Witten, \nupha{B460}(1996) 299.
\bibitem{Mar} E. Martinec, Phys. Lett. {\bf 367B} (1996) 91.
\bibitem{IM} H. Itoyama and A. Morozov, hep-th/9511126.
\bibitem{AN} Changhyn Ahn and Soonkeon Nam, hep-th/9603028 to appear in Phys. Lett. B.
\bibitem{RUS} A. Gorskii., A. Marshakov, A. Mironov, and A. Morozov,
hep-th/9603140.
\bibitem{Brz} T. Brzezinski, hep-th/9604101.
\bibitem{JP} See for example J. Polchinski, {\it String Duality-A Colloquium},  hep-th/9607050.
\bibitem{KV} S. Kachru and C. Vafa, \nupha{B450} (1995) 69.
\bibitem{MV}  C. Vafa, hep-th/9602022.
\bibitem{KKLMV} S. Kachru, et al., \nupha{B459} (1996) 537.
\bibitem{KHU} Changhyun Ahn, S. Hyun, and S. Nam, work in progress.
\bibitem{HO}  A. Hanany and Y. Oz, Nucl. Phys. {\bf B452} (1995) 283.
\bibitem{Marsha} A. Marshakov,  hep-th/9602005.
\bibitem{Koz} V.V. Kozlov, Russian Math. Surveys {\bf 38} (1983) 1.
\bibitem{BK} A.I. Bobenko and V.B. Kuznetsov, 
J. Phys. A : Math. Gen. {\bf 21} (1988) 1999.
\bibitem{BvM} C. Bechlivanidis and P. van Moerbeke, 
Comm. Math. Phys. {\bf 110} (1987) 317.
\bibitem{SE} W.H. Steeb and N. Euler, {\it ``Nonlinear Evolution
Equations and Painleve Test''}, World Scientific (1988), Singapore.
\bibitem{COFKM} P. Candelas, et al., \nupha{B416} (1994) 481.
\bibitem{HKTY} S. Hosono, et al., Comm. Math. Phys. {\bf 167} (1995) 301.  
\bibitem{KLMVW} A. Klemm, W. Lerche, P. Mayr, C. Vafa, and N. Warner, hep-th/9604034.
\bibitem{CFKM} P. Candelas, A. Font, S. Katz, and D. Morrison, \nupha{B429} (1994) 626.
\bibitem{BRS} A.I. Bobenko, et al., Comm. Math. Phys. {\bf 122} (1989) 321.
\bibitem{KT} V.B. Kuznetsov and A.V. Tsiganov,
J. Phys. {\bf A}: Math. Gen. {\bf 22} (1989) L73.
\bibitem{Kul} P.P. Kulish,  hep-th/9507070.
\bibitem{IY} K. Ito and S.K. Yang, \phlta{B366} (1996) 165.
\bibitem{BF} A. Bilal and F. Ferrari, hep-th/9605101.
\end{thebibliography}
\end{document}